# Two dimensional magnets: Forgotten history and recent progress towards spintronic applications

*David L. Cortie\*, Grace L. Causer, Kirrily C. Rule, Helmut Fritzsche, Wolfgang Kreuzpaintner, Frank Klose\*,*

Dr. David L. Cortie, Dr. Kirrily C. Rule
ARC Centre for Excellence in Future Low Energy Electronics Technologies

Dr. David L. Cortie, Dr. Grace L. Causer, Dr. Kirrily C. Rule
Institute for Superconducting and Electronic Materials, The University of Wollongong, Wollongong, NSW 2500, Australia

Dr. Kirrily C. Rule
Australian Nuclear Science and Technology Organisation, Lucas Heights, NSW 2234, Australia

Dr. Helmut Fritzsche
Canadian Nuclear Laboratories, Chalk River, Ontario, K0J 1J0, Canada

Prof. Frank Klose, Dr. Wolfgang Kreuzpaintner
Guangdong Technion-Israel Institute of Technology, Shantou 515063, P. R. China

*Abstract:* The recent discovery of two-dimensional magnetic order in van-der Waals materials has stimulated a renaissance in the field of atomically-thin magnets. This has led to promising demonstrations of spintronic functionality such as tunneling magnetoresistance. The frantic pace of this emerging research, however, has also led to some confusion surrounding the underlying phenomena of phase transitions in two-dimensional (2D) magnets. In fact, there is a rich history of experimental precedents beginning in the 1960s with quasi-2D bulk magnets and progressing to the 1980s using atomically-thin sheets of elemental metals. This review provides a holistic discussion of the current state of knowledge on the three distinct families of low-dimensional magnets: quasi-2D, ultra-thin films and van-der Waals crystals. It highlights the unique opportunities presented by the latest implementation in van-der Waals materials. By revisiting the fundamental insights from the field of low-dimensional magnetism, this review will highlight factors that can be used to enhance material performance. For example, the limits imposed on the critical temperature by the Mermin-Wagner theorem can be escaped in three separate ways: magnetocrystalline anisotropy, long range interactions and shape anisotropy. Several recent experimental reports of atomically-thin magnets with Curie temperatures above room temperature are highlighted.

## 1. Introduction

The theoretical premise of two-dimensional (2D) magnetism was firmly established at an early stage in 1944 when Onsager published his famous paper showing that a monolayer-thick Ising magnet has a phase transition to a long-range ordered state [1]. Experimental examples of ordered-2D magnets, however, have proven to be extremely rare in practice. Consequently, the important roles in technologies such as data storage [2], energy generation [3-4], water purification [5] and biomedicine [6] are all filled using traditional three-dimensional (3D) magnets based on bulk crystals, amorphous alloys or nanostructures consisting of dozens of atomic layers. With the recent discovery of a diverse library of 2D magnets [7], this trend may be poised to change. This review will highlight the opportunities presented by the new family of van-der Waals (vdW) magnets: two-dimensional atomic crystals featuring magnetic elements arranged in a lattice supporting intrinsic magnetic phase transitions. Within a narrow time-window of only two years, the ability to exfoliate vdW materials to form atomically-thin magnets has led to a succession of fundamental discoveries, and proof-of-principle spintronic functionality such as tunneling magnetoresistance, spin torque transfer and spin pumping devices. It is tempting to envision that future applications of these magnets may extend beyond pure spintronics into other realms, including chemical sensing and data storage. A critical prerequisite for most 2D or 3D magnetic applications, however, is the existence of a long-range ordered ferromagnetic state which emerges below a phase transition temperature ($T_c$). The spontaneous symmetry breaking below $T_c$ ultimately allows the material to retain a net magnetic moment in the absence of a magnetic field, which is central to many applications and forms the basis of all permanent magnets.

The magnetic order of 3D permanent "ferromagnets", such as "lodestone" ($Fe_3O_4$), was already exploited by the Ancient Greeks and Chinese thousands of years ago [8-9]. The concept of functionalizing 2D long-range order, however, was only envisioned in the past century. Onsager's solution was a landmark in theoretical physics and immediately had a large impact on the field of statistical mechanics and the study of phase transitions. However, it had a delayed impact in the field of experimental magnetism because it models a mathematically-simplified form of magnet that is difficult to observe in practice. The implicit simplification in the Ising model is that the magnetic moment on each atom has only one degree of freedom, and can point only up or down along one crystal axis (usually defined along ±z). In contrast, in most known magnets, magnetic moments behave like a semi-classical quantity and can be approximated using a 3D vector, particularly for systems with large moments (e.g. S=5/2 in $Fe^{3+}$) where quantum effects are diminished. It is also common to refer to the net atomic magnetic moments as a "spin" because the orbital contribution is typically small, and the moment arises entirely from the intrinsic spin of the electrons. If all three vector components (x,y,z) can play a role without any angular anisotropy, this is known as the classical Heisenberg model. Before 1950, the dominant view was that the best approximation to describe real magnets was the Heisenberg model, and in that context, the Ising model, and Onsager's solution, were regarded as theoretical simplifications with no detailed correspondence to experiment [10], or what experimental physicists term a "spherical cow model". However, by the late 1950s, a growing number of real magnets implementing Ising behavior had been discovered [10]. The quest to unveil new Ising magnets continues today because even 3D forms are relatively rare and only emerge from a special conspiracy of factors in the quantum chemistry and the local crystal structure, however, their unique levels of spin frustration often manifest in exotic magnetic states such as the spin ice state [11]. Since Ising magnets are the exception rather than the rule, most functional magnetic materials are still understood starting from the Heisenberg paradigm. In this context, a cornerstone in the physics of 2D magnets is the Mermin-Wagner theorem [12] which rigorously proved the absence of long-range order in 2D isotropic Heisenberg systems with short-range interactions owing to the thermal fluctuations, which are associated with divergent spin-wave dynamics ("magnons") in a magnetic state. Although the Mermin-Wagner theorem was published twenty years after the Onsager solution, it was widely misunderstood to suggest that magnetic order, and even crystal order, is impossible in two dimensions. However, a key point is that the latter theorem does not apply to Ising spin systems, or any spin system with significant anisotropy. Thus, the subsequent six decades offer several well-established examples of 2D and quasi-2D magnets.

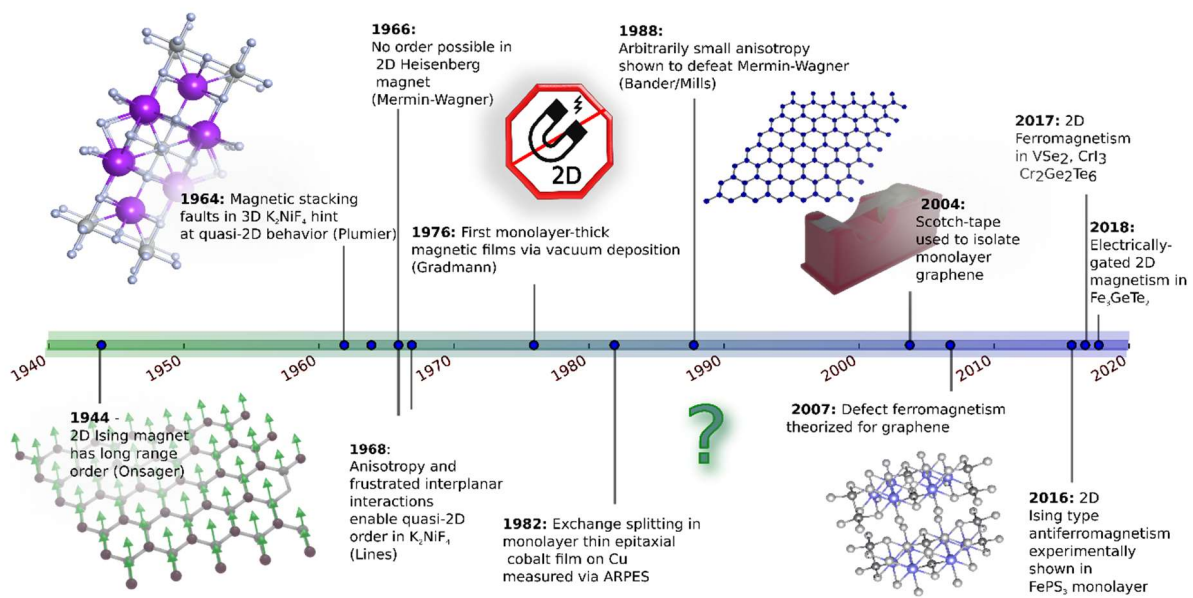

**Figure 1.** Timeline of selected important theoretical and experimental results in the field of 2D magnetism beginning with Onsager's solution in 1944, and culminating in the discovery of intrinsic 2D vdW materials in 2016.

**Figure 1** presents a time-line of significant events in 2D magnetism stretching from Onsager's solution to the present day. Three main directions within the field of 2D magnetism can be identified. Firstly, in the 1960s, quasi-2D magnets were identified in bulk crystals wherein weak-interplanar interactions prevent magnetic order in the third dimension [13-14]. This gave birth to a rich field of low-dimensional magnetism that continues to be active today. Secondly, parallel advances in thin-film growth during the 1970s to the 1990s led to the first atomically-thin film magnets [15], and enabled the first direct tests of Onsager's 2D model for magnetic monolayers [16]. The magnetism of ultrathin films and monolayers made from the classic transition metals has been reinvigorated in recent years using novel experimental and theoretical capabilities [17-18] [19] [20] [21].  The third and most recent implementation of 2D magnetism began shortly after the isolation of graphene in 2004 and has been progressing quickly since 2016 when the first 2D magnet based on an intrinsic vdW Ising compound was exfoliated [22].

While 2D magnetism is clearly not a new phenomenon, the purpose of this review article is to highlight the emerging opportunities arising from the discovery of the intrinsic vdW magnets. Prior to 2016, only a handful of atomically-thin magnets were known [15]. Although defect engineering of dilute magnetism in graphene and $MoS_2$ continues to attract interest [23-24], a seismic shift in the field has occurred towards a focus on intrinsic 2D magnets, particularly those exfoliated from crystals with 3D Ising order. The success of this approach is self-evident because it has unveiled an entirely new library of compounds, one which is rapidly expanding. The list of experimentally-verified examples includes $FePS_3$ [22] [25], $MnSe_2$ [26], $Fe_3GeTe_2$ [27-28], , $Cr_2Ge_2Te_6$ [29] [30] or $CrI_3$ [29] [31]. In the current era of 'fake news' and 'alternate-truths', it appears that scientific discourse itself is not immune from these trends, with several media releases making claims that the latter examples are the "the first time magnetism has been discovered in the 2D world of monolayers" [32], and arguing that vdW compounds are the "first truly 2D ferromagnets" [33] since "no 3D magnetic substance had previously retained its magnetic properties when thinned down to a single atomic sheet" [34]. Such claims are misleading given that a few examples of atomically-thin magnets have been known for decades. A prominent article also suggested that "physicists weren't even sure that 2D magnets were possible, until now" [34]. However, given the rigorous nature of Onsager's mathematical theory, and the many experimental precedents in the quasi-2D experiments, this statement implies the history of the field has been forgotten or overlooked. Nevertheless, beneath the veneer of 'media spin', there are profound scientific opportunities presented by the 'magnetic spin' systems based on vdW magnets. For the first time, the scientific community has access to a large, growing database of compounds with tunable properties that can be explored and functionalized with relative ease. In this task, a great deal can be learnt by revisiting the fundamental discoveries from the other historic sub-fields of 2D magnetism. While excellent reviews have been published on vdW magnets [7, 35] [36], they barely mention the insights gained from the non-vdW families of 2D magnets that predates them. Also, while there are several historic reviews and book chapters on quasi-2D magnetism [14] [37], there are fewer reviews on metallic monolayers [15, 38] with the most recent being the seminal review of ferromagnetic transition metal monolayers by Elmers in 1995 [15]. More generally, there appears to be no holistic account of the commonalities and overlap between these three different sub-fields of 2D magnetism. The goal of this review article is to summarize the current state-of-knowledge across the three research directions of 2D magnets and to explore and transfer insights.

This article is arranged into three sections as follows. **Section** 2 discusses the early work on quasi-2D magnetism in bulk compounds with weak-interlayer interactions. The latter provided the earliest hints that 2D magnetism could exist by showing the limitations of the Mermin-Wagner theorem as early as the 1960s. **Section 3** summarizes the study of atomically-thin metallic magnets that were enabled by breakthroughs in vacuum-deposition techniques developed in the 1970s to produce magnetic monolayers. **Section 4** summarizes the main results for the more recently discovered family of vdW magnetic monolayers and highlights the large library of potential candidate materials that remain to be explored.

## 2. Low-dimensional behavior in bulk 3D compounds

Magnetic order is analogous to crystallographic order from chemical bonding because it involves interactions between atoms along specific directions, and such interactions are often short-range in nature. For this reason, many mathematical solutions for magnetism such as the Ising model, are also applicable to certain types of crystal lattices, and vice versa. The basic ingredients for magnetic order are encapsulated in the model Hamiltonian where the total internal energy (H) is given by the summation:

$$\mathcal{H} = \sum_{i,j} J_{ij} S_i S_j$$

where $S_i$ ($S_j$) is the magnetic moment (spin) of the $i^{th}$ ($j^{th}$) atomic site. If the spins are confined to one axis such that $S=\pm Sz$, the Hamiltonian is identical to the Ising model. The exchange interaction $J_{ij}$ plays the role of the magnetic "bond" between atoms i and j. However, whereas chemical reactivity and strong bonding is ubiquitous for most elements across the periodic table (except the inert noble gases), magnetic spins and exchange interactions are much rarer in practice. In insulating materials, the spin moment depends on the number of unpaired electrons on the atoms in each oxidation state, and is often zero except for small number of elements, mostly in the 3d, 4f, 4d and 5d blocks of the periodic table. In iterant metallic magnets, the electronic bands can introduce additional complexity allowing delocalized moments (e.g. spin density waves), however, usually the main magnetic moment is still dominated by the unpaired spins existing around atoms[39]. Thus, given the rarity of local moments, magnetically inert materials without S=0 are very common across the periodic table, and most known compounds also do not exhibit magnetic order. Instead ,most materials are either diamagnetic or paramagnetic, resulting in a much weaker susceptibility that is impractical for most magnetic applications. Furthermore, magnetic ions sometimes find themselves in a frustrated environment from the point of view of their competing magnetic interactions due to the geometric arrangement of ions which is determined mostly by chemical factors on a higher energy scale. In fact, it is not uncommon to find magnetic systems where there is negligible interaction for two magnetic ions if they are far apart (>10 Å) ($J_{ij}=0$) or cancelling interactions due to mutually frustrated exchange pathways along some directions [14]. In this scenario, the magnetic order exhibits a lower level of dimensionality compared to the crystallographic one. For example, although a $K_2NiF_4$ crystal is 3D at the chemical level, the spin-bearing (S>0) sheets are decoupled along one direction, such that it is valid to think of each as an independent 2D thin film [40] [13] [41]. In the modern vernacular, this is the definition of a quasi-2D magnetic system. The magnetism in such low-dimensional materials shows varied and exotic behaviors due to the subtle balance between space dimensionality and spin dimensionality. Many exotic properties arise from the quantum nature of the Hamiltonian.

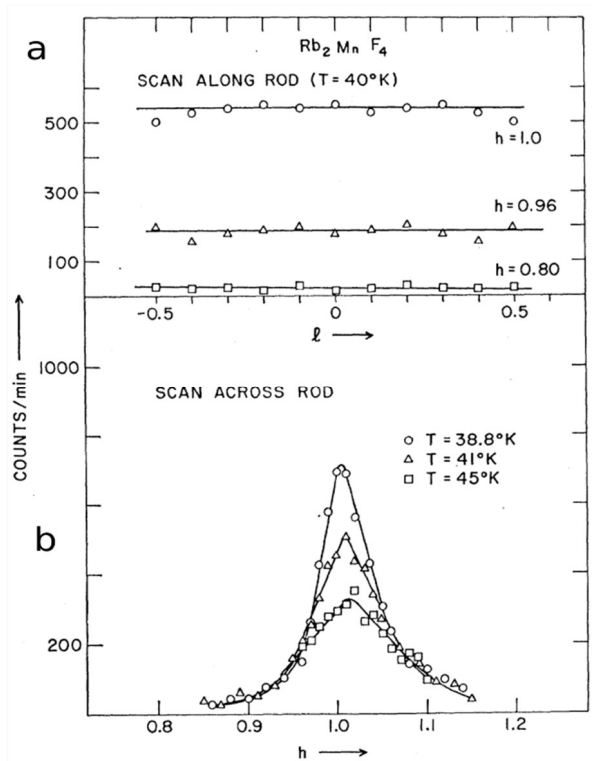

**Figure 2.** a) Neutron diffraction measurements for magnetic $Rb_2MnF_4$ show a Bragg ridge feature where no peaks are observed when scanning along the ridge, indicating no magnetic order in the third dimension. b) Clear peaks perpendicular to the ridge indicate magnetic order in the other two dimensions. (Reprinted from Ref. [43] with permission of the American Physical Society)

The first demonstrations of quasi-

2D systems were reported in the 1960s when experimentalists used the emerging technique of neutron diffraction to study magnetically-frustrated materials and vdW crystals for the first time [14] [40]. A key feature of neutron diffraction, that sets it apart from unpolarized X-ray or electron diffraction, is the high sensitivity of neutrons to magnetic structure. The interaction with magnetic moments (revealing the magnetic structure) is of the same order of magnitude as the nuclear interaction (revealing the crystal structure). Consequently, new Bragg peak intensities emerge corresponding to the magnetic sub-lattice ordering along any particular spherical direction [42]. A key finding in quasi-2D materials was that for certain crystal directions, magnetic ordering peaks could be observed, while for a direction perpendicular to the original axis, no peaks were observed, and instead so-called Bragg ridges appear, as exemplified in **Figure 2** for $Rb_2MnF_4$ [43]. This indicates the lack of long-range order along one direction and the presence of long-range correlations leading to long-range order within the other two perpendicular directions. With six decades of research behind it, it is now well-known that types of quasi-2D magnetism, and even quasi-1D magnetism, can be realized in bulk materials provided that the magnetic interactions in some directions are deactivated or suppressed by long-interatomic spacings or magnetic frustration. Today these systems continue to attract a great deal of interest because they are associated with very exotic states that can be activated by different combinations of these factors. Examples include spin-liquid states, Luttinger-liquid states [44] [45], spin-Peierls states [46], resonating valence bond states [47] as well as unconventionally ordered states via mechanisms such as order-by-disorder [48]. For example, high-temperature exotic superconductivity is one technologically useful phenomenon thought to be mediated by the low dimensional magnetic fluctuations [49].

A key insight from the field of low-dimensional magnetism in 3D crystals is recognizing that both the spin dimension ($n$) and the spatial dimension ($D$) are decisive factors to determine the magnetic ground state [37]. This provides an elegant classification scheme for all known magnetic combinations which can be used to quickly predict whether a given magnetic system will show long-range magnetic order, and this scheme is equally applicable to vdW magnets. The spatial dimension is the number of directions where non-zero magnetic interactions occur between spins in the systems. For example, most magnets are three-dimensional (D=3), however, $D$=2 occurs for atomically-thin monolayers or systems with zero-interplanar magnetic interaction. The spin dimension ($n$) relates to the degree of anisotropy in a system – that is, the number of axes that the spins are free to orient within. The latter factor is sometimes overlooked, but it is ultimately the decisive factor for the existence or non-existence of magnetic order. For $n$=1 the system is an Ising magnet. Planar anisotropy (XY) is described by $n$=2, while for $n$=3, the system is completely isotropic defining a Heisenberg system. **Figure 3** summarizes the classical scheme based on several decades of theoretical and experimental results in quasi-2D magnetism. The dark shaded regions indicate situations where order is theoretically impossible, whereas the lighter shaded regions indicate the potential existence of a long-range phase transition. It is worth noting that the entire column for *D*=2 is accessible using the new vdW magnets. While the Mermin-Wagner theorem [12] postulates the absence of long-range order in 2D isotropic Heisenberg systems (i.e. for *D*=2, *n*=3), it places no such restrictions on *D*=2, *n*=1, where order is possible in two dimensions. A particularly interesting case occurs for *D*=2, *n*=2, because a true XY magnet should not exhibit true long-range order, but may instead undergo a more exotic type of phase transition proposed by Kosterlitz-Thoules [50]. In this regime, at any non-zero temperature, the spins are unable to align perfectly along a common direction and instead form so-called 'topological' defects, which nonetheless still support a net magnetization. The magnetization behavior near this unusual phase transition is entirely different from a standard phase transition because the magnetization does not go to zero at the critical temperature [50]. Although a 2D magnet with XY anisotropy should, in theory, be the ideal setting to observe this exotic transition, in practice, to the best of our knowledge there are no experimental realizations of this using ferromagnets, and the experimental examples appear mostly in superconducting systems. In general, shape anisotropy occurring in ferromagnets flakes and thin-film magnets may ultimately prevent a perfect XY Hamiltonian [15], however recently suppressed order was reported in ultra-thin forms of the XY-type antiferromagnet $NiPS_3$ [51]. Finally, although no order appears to be possible in one spatial dimension for any spin dimension, an important caveat is that Figure 3 assumes the standard picture of short-range interactions in the magnet. For instance, it is known that long-range interactions can induce, all *n* and *D* combinations have the potential to show magnetic order. For example, long range magnetic potentials can induce critical phase transitions in 2D magnets where they would otherwise be forbidden [52] and similar arguments can even induce order in 1D [53]. Nevertheless, the existence of sufficiently long-range reactions is unproven experimentally, and probably difficult to realize in practice because most magnetic exchange interactions are either short-ranged or relatively weak. For example, even the "long-range" Ruderman–Kittel–Kasuya–Yosida interaction in metals is confined to several nanometers and does not fulfill the criteria to induce order in a 2D Heisenberg. Another nuance is that the spin anisotropy that defines *n* actually lies on a spectrum of values, and the simplified picture of Ising, XY and Heisenberg are only useful limits. In most magnets, the situation lies somewhere in between these extremes.

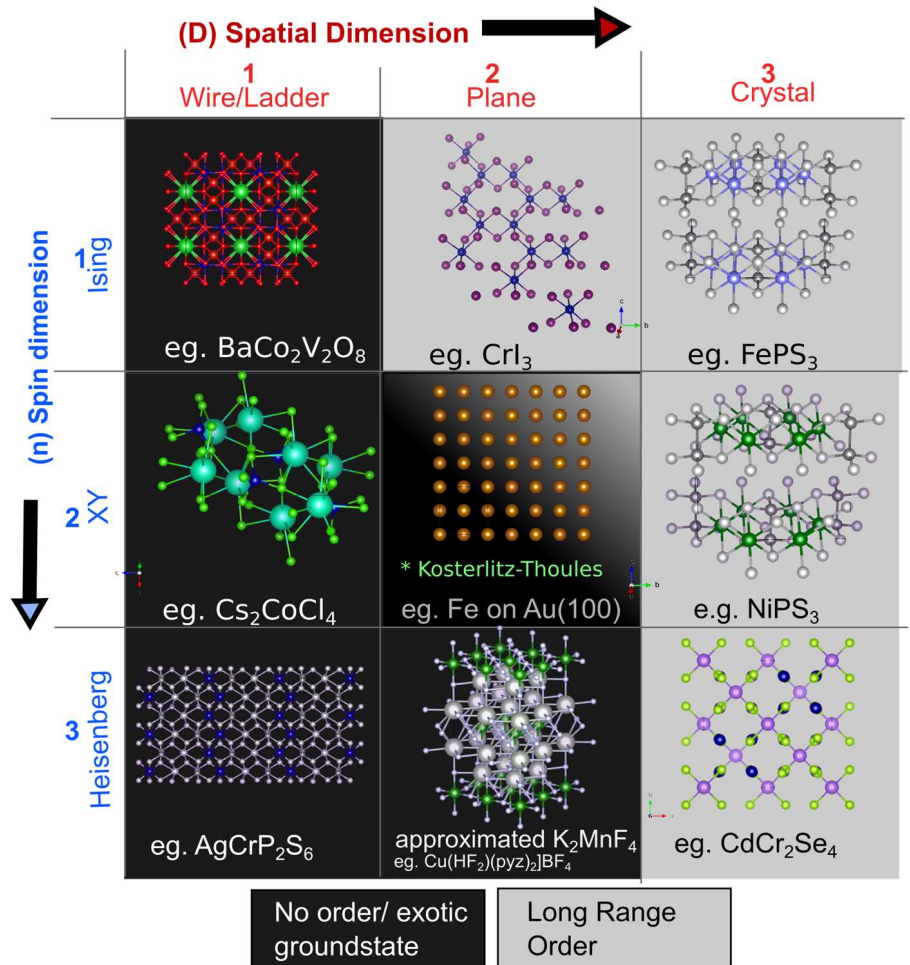

**Figure 3.** Classification scheme based on the real-space dimension (D) and the spin dimension (n), with examples of materials studied experimentally and their crystal structures. Although in some cases the crystals are 3D substances, the reduced dimensionality (D) arises because magnetic interactions are zero in some directions. Black backgrounds indicate conditions which strictly exhibit "no order", whereas white squares indicate long-range order. The $D$=2, $n$=2 is a special intermediate case with no long-range order but short-range order via a Kosterlitz-Thoules phase transition.

An important diagnostic tool in the experimental study of magnetic phase transitions is the critical exponent for magnetization (*β*) as this is also an indication of the spin dimension [39]. By measuring the behavior of the magnetic moment *M(T)* as a function of temperature near the magnetic phase transition, a relationship emerges which allows *β* to be determined:

$$M(T) = const \left(\frac{T_C - T}{T_C}\right)^\beta$$

The critical exponent β is a fingerprint of both the spin and spatial dimensionality. For instance β = 1/8 for a 2D Ising magnet on a square lattice, whereas β = $3\pi^2$/128 = 0.23 for 2D-XY behaviour [15]. In bulk 3D compounds, the D is fixed by the innate quantum chemistry of the crystal which cannot be altered easily. Intuitively one may expect that many vdW materials with large interlayer gaps would also exhibit quasi-2D magnetism, however, this is not in fact the case. An important general trend is that the interactions are not negligible in most vdW materials because significant electronic exchange interactions occur across the vdW gaps (~ 6 Å), and so 3D vdW systems usually exhibit true 3D order[13-14]. Indeed, this was realized at a very early stage, and in a very early review article by Lines [13-14]. Here he showed examples of how well-behaved quasi-2D order arises from frustrated systems which produce competing exchange pathways, such as in $K_2NiF_4$. However, the ability to exfoliate vdW crystals allows a facile method to gain access to the (D-1) state. The implication is that there is a much wider range of atomically-thin vdW magnets than previously thought, because even when D=3 in the bulk, these may well exhibit 2D magnetism through magnetically isolated atomic layers. This feature has been established for $Fe_3GeTe_6$ and $Cr_2Ge_2Te_6$, both which exhibit 3D magnetism in bulk (D=3), however, they can be exfoliated to form D=2 layers [7, 30]. Therefore, it is not necessary to use a strictly quasi-2D crystal to create an atomically thin layer, in fact, a nearly-2D (n2D) crystal with some degree of interplanar interactions is an equally good candidate for exfoliation.

### 3. The first atomically-thin magnets in elemental Co, Fe and Ni

After the discovery of the quasi-2D bulk compounds, and with the development of ultrahigh-vacuum technologies for the growth of thin epitaxial films, it became clear that there was no major barrier to achieving long-range order in monolayer thin materials. Gradmann was one of the first in the 1960s to report ferromagnetism in ultrathin films [54-55]. He was able to measure the temperature dependence of a 1.8 monolayer (ML) thick NiFe alloy film prepared on Cu(111) and determined a Curie temperature of about 220 K for this thickness regime. Surprisingly, however, interest in the area of ultrathin film magnetism only started to expand in the early 1980s. A key result was the angle-resolved photoemission experiment which showed exchange splitting in a monolayer of cobalt[56]. More generally, ultrathin films began to attract increased attention in this period once it was realized that they exhibit various unusual magnetic properties such as spontaneous magnetization perpendicular to the film surface, altered Curie temperatures, ferromagnetism or ferromagnetism induced interfacial magnetic moments in adjacent paramagnetic or diamagnetic films or degraded ferromagnetism in surfaces or near surface layers. Such effects are not observed in bulk samples. These new magnetic properties which manifest on the nanometer length scale, provide both fundamental questions for basic research and new opportunities for technology. The research in thin magnetic films culminated in the discovery of oscillating interlayer coupling and giant magnetoresistance [57]. This was the basis of awarding the 2007 Nobel Prize in physics to P. Grünberg and A. Fert in 2007, which rejuvenated the field [58-59]. Today, thin films with a thickness of tens of monolayers are the basic units in magnetic recording, and the discovery of giant magnetoresistance led to a drastic increase in the sensitivity of computer hard disk read-heads - the first examples of spintronic devices - enabling a staggering increase in the areal density of magnetic recording devices. In the past decade, magnetic tunnel junctions have been increasingly deployed in magnetic ram (MRAM) developed to industrial-scale to enable non-volatile memory [60] by exploiting the reciprocal relationship between spin transfer torque and tunnelling magnetoresistance [61].

Beginning in the 1980s, there were a couple of interesting questions that attracted a lot of attention in the scientific community and are still being explored and rediscovered today:
1) How thin can a material become and still retain magnetic order (e.g. ferromagnetism)?
2) How can a magnetic layer be classified as a true 2D object?
3) What is the critical behaviour of the magnetization and fluctuations in thin films close to the Curie temperature $T_C$?
4) How does the magnetic anisotropy of a thin magnetic film change due to the interaction with a substrate or a cover layer?
5) What is the reason for a spin reorientation transition (SRT), i.e. the phase transition from in-plane to perpendicular magnetization?

6) Is the film in the region of the SRT still ferromagnetic?

During the period, 1980-2000, ferromagnetism of Fe, Co, and Ni monolayers was experimentally proven for various substrates and cover layers (see Table 1). In addition to the phase transition temperatures, the available methods also allowed the determination of the critical exponent β as introduced in Section I. According to Qiu et al. [62], thin films should be categorized in two classes: films with uniaxial anisotropy should follow the 2D-Ising model [63] with a critical exponent of β=1/8, whereas films without uniaxial anisotropy should show a 2D-XY behavior with β=$3\pi^2/128$=0.23. Indeed, a review by Elmers [15] shows that the experimentally determined critical exponents of ultrathin films are in agreement either with the Ising model or the 2D-XY model. The observation of order in the 2D-XY magnet, Fe on Au(100), was in contrast to the predictions of the Kosterlitz-Thoules state proposed for this system (Figure 3) for $D$=2, $n$=2. Thus, already at an early stage, the applicability of the Mermin-Wagner theorem [12] was discussed, and it was recognized that order in 2D-systems must be triggered by anisotropies [63], contrary to 3D systems where it can exist without anisotropies. A key insight from this era is that all thin films possess strong magnetic anisotropies, arising primarily from the shape anisotropy and the magnetic surface anisotropy (MSA) of the film which was predicted by L. Néel [64] as a result of the break of the local symmetry at the thin-film interfaces. Thus in practice, the Mermin-Wagner theorem is usually defeated by these additional factors, even when the underlying system is not Ising-like. An interesting situation arises when the MSA supports perpendicular magnetization and may overcome the shape anisotropy at small thicknesses leading to perpendicular magnetization. Such a SRT has been observed in many systems, e.g. Cu(111)/NiFe/Cu [65], Au(111)/Fe [66], Ag(100)/Fe [67], Cu(100)/Fe [68], Cr(110)/Fe [69], Pd(111)/Co/Pd [69], or Au(111)/Co/Au [70]. These SRTs have been observed as a function of film thickness as well as temperature. Interestingly, the character of the SRT is determined by higher order anisotropy terms and three types of SRT can be distinguished [69]: (i) a continuous rotation from in-plane to perpendicular orientation, (ii) a co-existence of in-plane and perpendicular magnetization, (iii) the special case of zero anisotropy at the SRT. The continuous rotation with a canted spin structure has been observed in Au(111)/Co/Au layers [71], the discontinuous SRT in Pd(111)/Co/Pd can be concluded from the measured anisotropies. The special case of zero anisotropy and the associated loss of long-range order according to the Mermin-Wagner theorem is highly unlikely to be observed experimentally because it would require an exact compensation of all higher-order magnetic anisotropies at the critical thickness of the SRT.

Table 1 summarizes past reports of monolayer magnets based on Fe, Co and Ni. It is important to note that some of the experimental results reported by different groups in Table 1 do not agree. This is not too surprising since all measurements in the monolayer regime are extremely challenging: the magnetic signal from the monolayer is extremely small, and the preparation of an ideally flat monolayer requires a particular procedure that has to be followed carefully by the experimentalists. There are a couple of issues that can contribute to discrepancies:
  i) The structural quality of the substrate (crystallinity, roughness, chemical purity)
  ii) The substrate temperature during the deposition
  iii) General vacuum condition and the presence of residual gases during the film preparation
  iv) Film thickness calibration

**Table 1.** Examples of experimentally-validated ferromagnetic Fe, Co, and Ni monolayers on various substrates, covered and uncovered. The experimental methods are: SPPE (Spin-polarized Photo Emission), ECS (Electron Capture Spectroscopy), MOKE (Magneto-optical Kerr Effect), SPLEED (Spin-polarized Low Energy Electron Diffraction), SQUID (Superconducting Quantum Interference Device), CEMS (Conversion Electron Mössbauer Spectroscopy), TOM (Torsion Oscillation Magnetometry), BLS (Brillouin Light Scattering), PNR (Polarized Neutron Reflectometry)

| System | $T_c$ | $\beta$ | Method | Year published | References |
|---|---|---|---|---|---|
| Ag(100)/Fe | >30 K | | SPPE | 1987 | [72] |
| | <290 K | | ECS | 1989 | [73] |
| Ag(111)Fe | 80 K | 0.137 | MOKE | 1991 | [74] |
| Au(100)/Fe | 315 K | 0.22 | SPLEED | 1989 | [75] |
| Au(111)/Fe/Au | >10 K | | SQUID | 1989 | [66] |
| W(110)/Fe | 210 K | | CEMS | 1989 | [76] |
| | 230 K | 0.123 | SPLEED | 1994 | [77] |
| W(110)/Fe/Ag | 296 K | | CEMS | 1987 | [76] |
| | 282 K | | TOM | 1989 | [78] |
| Cu(100)/Co | 290 K | | MOKE | 1992 | [79] |
| | <290 K | | BLS | 1992 | [80] |
| Cu(111)/Co/Cu | 433 K | 0.125 | TOM | 1992 | [16] |
| Ag(100)/Co/Ag | >5K | | PNR | 1992 | [81] |
| Cu(111)/Ni | 160 K | | MOKE | 1990 | [82] |
| Ag(100)/Ni | >110K | | MOKE | 1989 | [83] |

By the 1990s, better convergence was found and a range of new techniques were applied, such as scanning tunneling microscopy. A major achievement was the fascinating agreement of the experimental data with the exact solution of the Ising model in the Cu(111)/Co/Cu system [16], as shown in **Figure 4 b)**. This provides a strong argument that ferromagnetic monolayers are true 2D-magnets. The experimental study of ultra-thin metallic films continues to be an active topic of research to this day. In contrast to vdW magnets, which are typically small flakes (1 µm), it is possible to synthesize large-area monolayer magnets. Recently, even electrostatically switchable magnetism has been induced on the surface of Pt thin films by applying ionic liquids as the gating media[84]. Along these lines of interface-induced magnetism, a promising development in recent years has been deploying molecules (e.g. $C_{60}$) to induce magnetism in non-magnetic metals (e.g. copper), leading to the idea of the "spinterface" [85] [86]. Yet another promising development has been to interface ferroelectric or magnetoelectric materials with traditional magnets (eg. Fe/$BaTiO_3$ [87] or Cr/$Cr_2O_3$ [88]) to enable electric-field switching and robust control over exchange bias. Consequently, the magnetism of ultrathin films and monolayers made from the classic anti- and ferromagnetic transition metals remains a vibrant field and is currently being reinvestigated with novel experimental and theoretical capabilities [17, 19] [20] [21]. The crystalline order of metallic monolayers is now widely studied using scanning tunneling microscopy (**Figure 4 a**). Since the 1980s an important technique to characterize such samples came into play: polarized neutron reflectometry (PNR). It was recently used to study the growth of Fe on Cu(001)/Si(001) (**Figure 4c**) from the sub-monolayer regime where island growth dominates (**Figure 4d**) to a final thickness of approximately 30 monolayers [89].

The key lessons learnt from the field of ultra-thin metallic films, equally applicable to vdW magnets, relate to the importance of the demagnetization fields and edge-anisotropy arising both from the finite-thickness and potentially, the island-like nature of the monolayer in the lateral dimension. Simple models such as the Heisenberg/Ising model do not consider the role of the classical electromagnetic dipolar field because its energy is so low compared to the exchange interactions. However, while the anisotropy is small, it is enough to modify the conclusion of the Mermin-Wagner theorem since it represents a deviation from the ideal Heisenberg magnet. In bulk quasi-2D magnets this anisotropy may be disregarded, however, in 2D magnets or 2D island magnets this is important because the shape anisotropy energy of ultra-thin films and other flat nanomagnets is strong and severe. Elmers gave a detailed discussion of the effect of finite-lateral size and superparamagnetic effects in island-monolayers, and it appears that this can in some cases stabilizes order in situations which would normally be forbidden (e.g. in the XY 2D magnet for $D=2$, $n=2$) [15]. An important point is that standard density functional theory (DFT) calculations widely used in modern theoretical work related to 2D magnets do not model the classical dipolar anisotropy, nor consider the effect of nanoscale edges on the magnetic anisotropy. However, doing so may be critical because some anisotropy is a necessity for 2D magnetism, and in many cases, this may arise solely from the shape anisotropy term. It is also possible to envision cases where the opposite is true, and the dipolar anisotropy cancels out with the intrinsic anisotropy to yield a magnet that cannot order. The proper treatment of dipolar and spin- orbit anisotropies remains an emerging field with DFT and one that has barely been explored within the context of the vdW magnets.

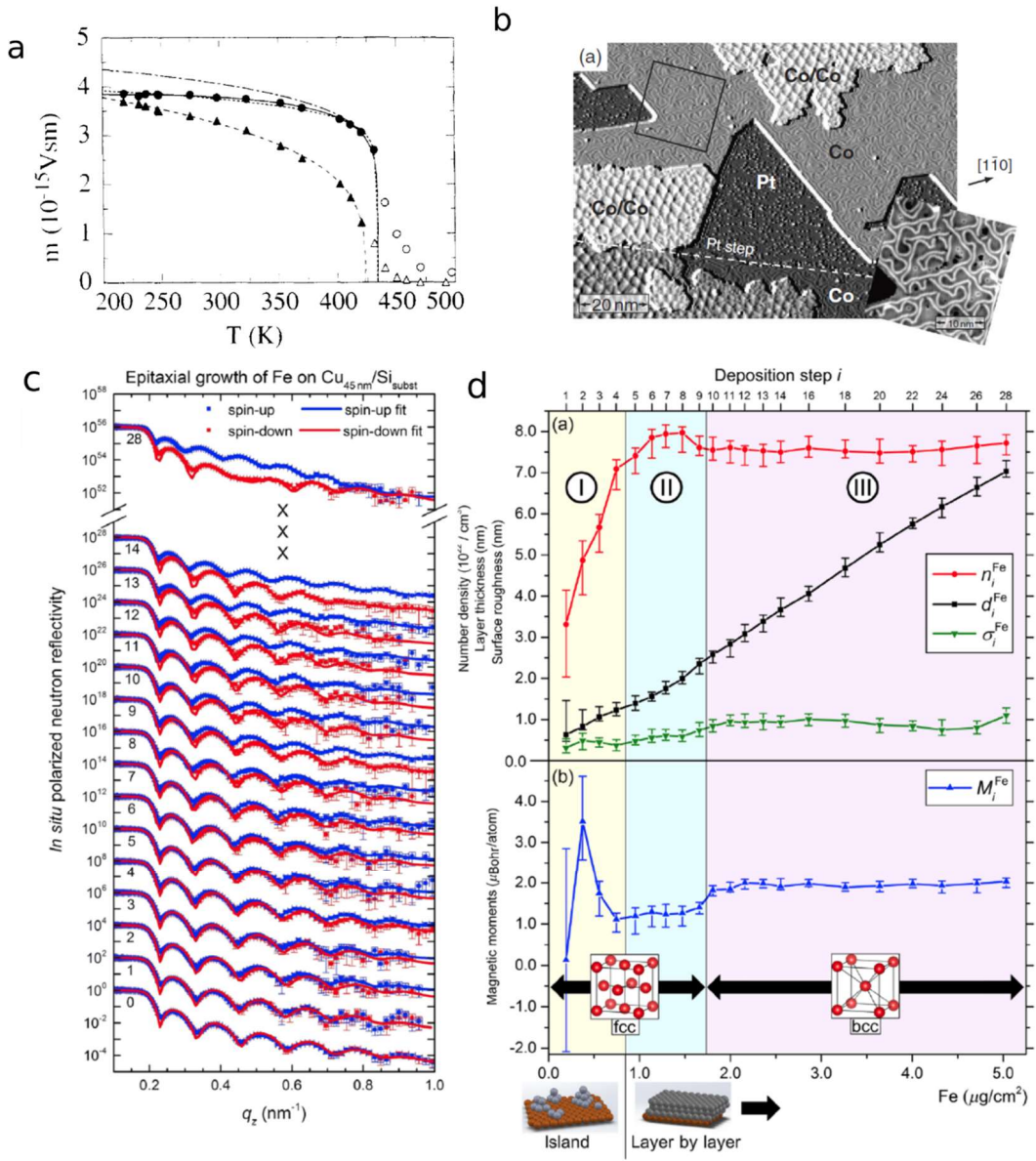

**Figure 4. a**) Perfect agreement of the magnetization of a cobalt monolayer with the 2D Ising model from Ref. [16] **b**) Scanning tunneling microscope image of cobalt monolayer and bilayer on Pt substrate from Ref [20]. **c**) Polarized neutron reflectometry patterns for epitaxially *in situ* grown Fe on Cu(001)/Si(001). Each pair of curves is vertically shifted by 2 orders of magnitude for better visibility. The number below the regime of total reflection (0 to 28) denotes the deposition step i. In each deposition step the equivalent amount of Fe, required to form one atomic layer was deposited. **d**) Summary of the information extracted from the polarized neutron reflectometry showing the development of the structural (a) and magnetic (b) thin film parameters (vic. number density $n_{Fe}$, film thickness $d_{Fe}$, interfacial roughness $\sigma_{Fe}$, and magnetization $M_{Fe}$). Three main regimes (I–III) with distinctively different characteristic behaviors can be identified. Shown are also the concluded growth modes (island and layer-by-layer) and crystalline structures (face-centered cubic and body-centered cubic). (Reprinted from Ref. [89] with the permission of the American Physical Society)

## 4. Third-generation 2D magnets: vdW magnets

The discovery by Geim in 2004 [90] that 2D graphene could be exfoliated from graphite simply using 'scotch' tape was a watershed moment in modern material science and gave birth to a vibrant new field of 2D materials. Although graphene itself was not entirely new (Geim et al. credited earlier work beginning in the 1960s [91]), the new capability to achieve free-standing monolayers electrically-isolated from their substrate was intensely attractive for real-world functionality. For example, this soon led to the demonstration of the Quantum Hall Effect [92], extending up to room temperature in the presence of high magnetic fields (~ 45 T) [93]. This profound discovery has inspired a new theme in modern material science aimed at developing dissipationless electronics that function at high temperature. This has ultimately led towards a recent demonstration of the Quantum Anomalous Hall Effect using thin films of magnetic topological insulators operating in zero field (*83*). The current family of 2D materials has grown rapidly in the past decade, and now includes a variety of metals (e.g. $NbSe_2$) [94], semimetals (e.g. graphene) [90], semiconductors (e.g., $MoS_2$) [95], insulators (e.g. hexagonal boron nitride) [96], superconductors (e.g. $NbSe_2$) [97] (*84*) and topological insulators (e.g. $WTe_2$) [97]. Although this list is an impressive set of building blocks for technology, it was incomplete until recently because notably absent from this list was a robust 2D magnet. For this reason, beginning with the discovery of graphene in 2004, there has been a quest to find two-dimensional vdW materials with robust magnetic properties and high Curie temperatures.

Intensive work explored the notion of engineering or "imprinting" magnetic order into graphene using a variety of different methods. It has been reported that graphene can be made magnetic by ion implantation or intercalation of transition metal atoms [98] [99]. Alternatively, magnetic spins can also be formed via hydrogenation [100], by strain [101], by proximity effects [102] or by the introduction of point defects [103]. A very comprehensive review on the topic has recently been published in Ref. [24], and here we just highlight a few points. Like other carbon allotropes, the flexible $sp_x$ bonding schemes available to the carbon atoms generally make it unfavorable to form unpaired electrons, and consequently, the carbon atom is usually magnetically inert (S=0) except in the presence of defects. Recently, however, there has been an experimental report of H-graphene on a SiC substrate exhibiting ferromagnetism, lending hope that graphene may be converted to a useful magnet [104]. However, the role of the substrate-graphene interactions appear to be pivotal in this case, and it is unclear whether the graphene monolayer itself can be intrinsically magnetic. Meanwhile, most proposed schemes focus on controlling the concentration of the defects, as well as the interatomic spacing/absorption location, which would be extremely difficult in practice. Currently, defect-engineering remains one of the most challenging aspects of material science, particularly when magnetism is involved[105-106]. Ion implantation of magnetic ions has also been reported for other 2D materials [107].

One avenue to avoid the complexities of defect-engineering, and to obtain reproducible, high magnetizations, is to study robust intrinsic magnets comprised of a dense lattice of magnetically active (S>0) atoms. In this area, a key development was the exfoliation of $FePS_3$ which consists of Fe moments distributed onto a honeycomb lattice. Past work on the bulk $FePS_3$ compound had previously shown that the bulk 3D compound was a $D=3$, $n=1$ (Ising model) [108] [108]. This lent hope to the idea that antiferromagnetism would survive in two dimensions to give a D=2, n=1 system. Subsequent Raman scattering by two separate groups in 2016 on 2D $FePS_3$ confirmed antiferromagnetism, which provided incontrovertible evidence of an intrinsic magnetic phase transition [22] [25]. **Figure 5** presents the Raman scattering near the phase transition temperature for different numbers of $FePS_3$ monolayers. This provided a major conceptual advancement since it showed a path by which 2D magnetism was indeed possible in vdW compounds. However, $FePS_3$ is antiferromagnetic and consists of two sub-lattices that are antiparallel giving negligible net magnetization. Therefore, it was highly desirable to locate a ferromagnet for most applications. It has long-been known that the ferromagnetic and antiferromagnetic versions of the Ising model are effectively identical mathematically and share the same phase transition. Thus, it became obvious that a ferromagnetic Ising material would be a good starting point to locate a 2D ferromagnet.

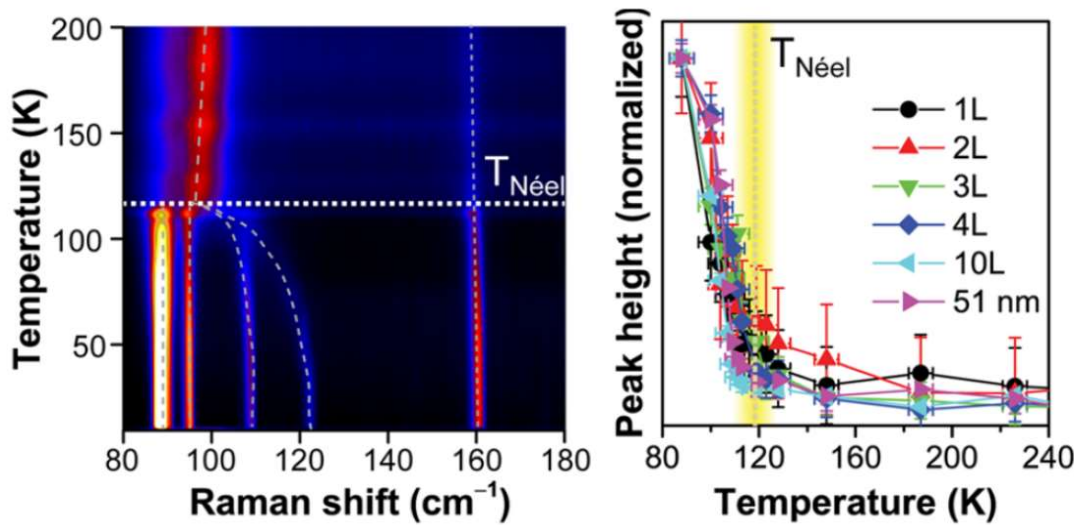

**Figure 5.** a) The splitting of the Raman peak at the Néel temperature (i.e. 110 K) of monolayer $FePS_3$ provides incontrovertible evidence of antiferromagnetic order in 2D. b) Multiple exfoliations allow the study of the thickness dependence of the ordering temperature, and the effects are weak for the $FePS_3$ system because of the innate Ising behavior and quasi-2D nature of the bulk crystal. (Reprinted from Ref. 22 with the permission of the American Chemical Society)

Within a year in 2017, Gong et al. showed that $Cr_2Ge_2Te_6$ was a ferromagnetic insulator [30]. In tandem $CrI_3$, was reported by Huang et al. as another example of an intrinsic 2D ferromagnet which shows striking layer-dependent ferromagnetism [29] [31]. Whereas the monolayer form of $CrI_3$ shows an observed remnant magnetization in a single layer indicating ferromagnetism, this is absent in a bilayer, indicating antiferromagnetism [29] i.e., the two layers having oppositely oriented spins. This is consistent with metamagnetic behavior long-known in this compound where subsequent sheets have an antiparallel spin structure. Interestingly, the net magnetization is again recovered in a trilayer. This indicates the strong layer-dependent interplay between different mechanisms that stabilize magnetic ordering. The magnetism in CrI$_3$ is particularly robust and it has already been incorporated into several functional devices. Important first steps towards technological applicability of 2D vdW magnets was demonstrated by developing device fabrication techniques [27] [109], and by detecting the presence of magnetism via electron tunneling [110] and in turn, influencing the magnetism in vdW crystals using electric fields [31] [111] [112]. In this context, e.g. a tunneling magnetoresistance device prototype, utilizing CrI$_3$ as the tunnel barrier in heterostructures, has been realized [109]. However, both CrI3 (T$_c$<45 K)and Cr$_2$Ge$_2$Te$_6$ (T$_c$<40 K) have relatively low temperature phase transitions.

A candidate with higher temperature was Fe$_3$GeTe$_2$, given that the bulk compound orders at T$_C$~220 K, however initial studies showed that the magnetic order temperature was strongly suppressed for thinner flakes [27-28], and indeed it appears T$_c$ in monolayer flakes is less than 10 K (27). This strong thickness dependency was quite different from that of FePS$_3$ and CrI$_3$, indicating that thermal fluctuations (eg. the Mermin-Wagner theorem) are indeed crucially important in some cases. Recent work, however, showed that ionic or electric gating could stabilize these properties to above room temperature, which is extremely promising [27]. Figure 6 shows the fabrication methods deployed by Deng et al. to produce electrically-gated devices[27]. This process also allows the possibility to study their magnetism as a function of atomic layer thickness.

Room temperature ferromagnetism in 2D vdW materials could open the possibilities for a wide range of devices and functionalities. The initial reports of CrI$_3$, Cr$_2$Ge$_2$Te$_6$ and Fe$_3$GeTe$_2$ showed relatively low Curie temperatures. While the latter could be enhanced above room temperature with electric fields, there are benefits in a system with an intrinsically high temperature capability. In 2018, two other room-temperature ordered vdW magnetic compounds, MnSe$_2$ and VSe$_2$, [26] [113] were reported. As these materials are both ordered intrinsically at room temperature, they offer a promising opportunity for spintronic applications. However, it is important to note that in three dimensions neither of the parent materials are magnetic, so the magnetism at the 2D limit is of a different origin. There is still a strong contemporary debate around the origin of 2D magnetism in VSe$_2$, particularly surrounding the role of the substrate-film interactions [35], and it is of great interest to see if it can be widely reproduced and utilized by other groups. Recently, a substantial number of additional 2D ferromagnets have been predicted by theory, see Ref. [35]. Systematic theoretical screening for vdW crystals with possibly the highest Curie temperatures has recently been published [114] to identify new candidates for monolayer magnetism. The highest transition temperatures exceeding 1000 K have been predicted in so-called 'Mxene' compounds, having the formula M$_2$AX$_2$ where M is a metal (Mn, Cr, Ti), A=(C, N) and X =(O, OH, F, Cl). However, the experimental validations of these predictions have not yet appeared, and more generally, techniques to reliably predict complex forms of 2D magnetism are still in their infancy.

In addition to magnetoresistance, other spintronic functionalities such as spin-pumping to convert charge-currrent and spin current have been demonstrated using 2D materials, for example, spin current has been injected into graphene using a ytriium iron garnet layer[115]. Spin torque transfer has also been demonstrated using MoS$_2$ on permalloy [25]. In this context, all-electrical control has also been demonstrated in bilayer CrI$_3$ below 40 K [111]. These three phenomena are the key ingredients in a number of devices including magnonic processors, magnon transistors and spin-torque junctions for non-volatile storage. However, it should be noted that MoS$_2$ and graphene are themselves non-magnetic, and the magnetic functionality in the latter devices relies on incorporating then with a traditional 3D magnet (eg permalloy).The grand challenge in this field is to demonstrate similar effects using a high temperature 2D magnet.

The experimental techniques used to investigate 2D vdW magnetism and isolate promising materials are still much less mature than the techniques used for quasi-2D or metallic thin-film monolayers because the latter have been investigated for decades. New techniques will be needed to investigate the more complex magnetic forms. Samples for investigation of 2D magnetism in vdW materials are mostly not produced as large area thin films, e.g. by vdW epitaxy [26],[116],[117] but typically as small micrometer-scale flakes by exfoliation [28]. As a consequence, the classic macroscopic analysis methods like superconducting quantum interference device (SQUID) magnetometry [118] cannot generally be applied for the analysis of vdW monolayer magnets without adaption. For example, neutron diffraction experiments generally require large masses, meaning that past thin-film work has been restricted to centimeter-wide crystalline films with thicknesses in the range of 100 – 400 nm [119] [120]. Instead, local probes or probes that can be focused onto certain sample areas of interest have an advantage, for example, Scanning Tunneling Microscopy (STM) [99],[121]. These can be combined with microscopy, like magneto-optical Kerr effect (MOKE) [29] [122] [111] and magnetic-circular dichroism (MCD) [111] are applied. Raman spectroscopy, using focused or confocal beams is also promising in that it can reveal details of complex magnetic order parameters, for example, antiferromagnetism in small samples [22, 25] [123-124]. Progress in the field of micro-spin angle resolved photoemission spectroscopy (ARPES) has enabled a technique to measure spin-resolved band structures for small samples [125]. A SQUID-based magnetic microscope [126] has been used to investigate the room temperature magnetism of $MnSe_2$ films [26]. However, an outstanding issue for these types of techniques is correctly assessing the importance of magnetic impurities. An extensive overview of the operational principles of most of the techniques is given in the review of [7]. Two additional techniques to study magnetic interfaces using quantum beam science are low energy μSR [127] and the related β-NMR method[126]. Both of these techniques monitor the polarization of highly spin-polarized particles implanted near the surface, thereby offering sensitivity to small magnetic moments and fluctuating magnetic fields nears surfaces[128]. Past work has demonstrated sensitivity even to paramagnetic magnetic monolayers sensed via dipolar coupling to a distance of several nanometers [129]. Ultimately, advanced techniques will be needed to study more exotic magnetic structures, not simply ferromagnetism, but also cycloidal antiferromagnetism and skyrmionic states.

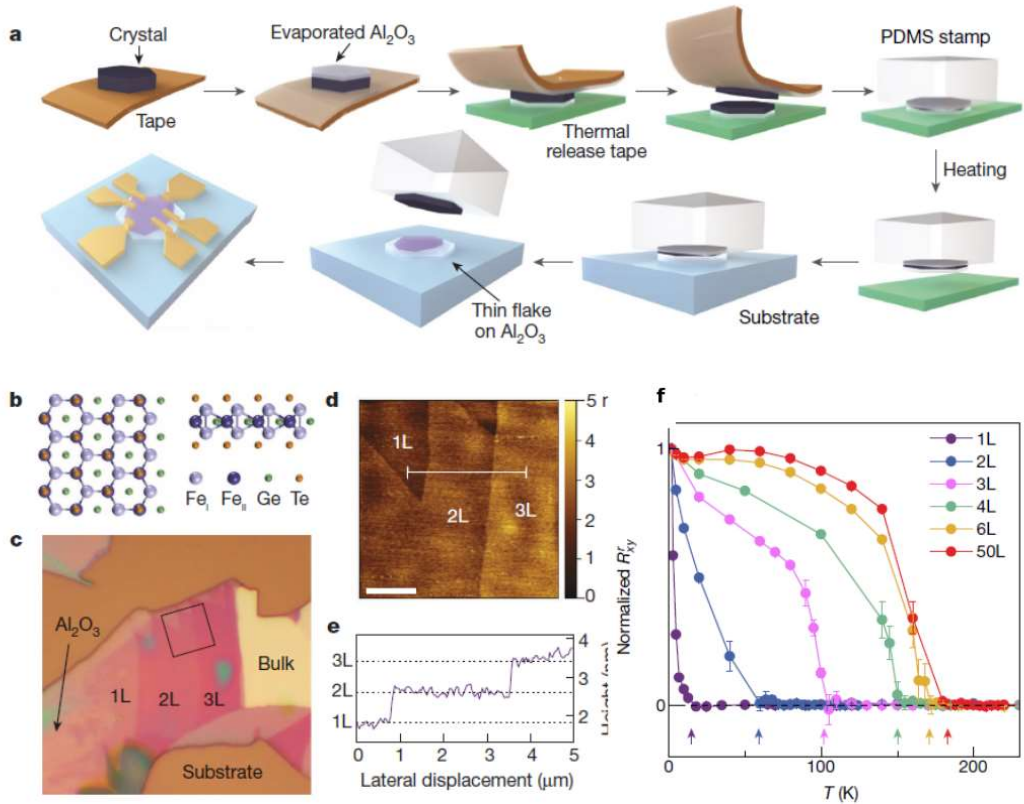

**Figure 6.** a) Schematic showing the fabrication process for $Fe_3GeTe_2$ electric-field device. b) Crystal structure of $Fe_3GeTe_3$ along the c-axis (left) and in the basal plane (right) c) Optical micrograph showing regions of the monolayer material d) Crystal structure of $Fe_3GeTe_3$ perpendicular to the c-axis d) Atomic force microscopy image of the monolayer regions in the sample. e) Atomic force microscopy image of layer height f) Thickness-dependent magnetization in $Fe_3GeTe_3$. Reprinted from Ref. 27 with the permission of the Nature Springer publishing group.

The field of vdW magnets is rapidly expanding and many more novel compounds are expected to be discovered. **Figure 7** shows a diagram of experimental examples of magnetic states in vdW compounds. There are many attractive and relatively unexplored topics in vdW magnets. Theoretical calculations to predict better candidates are critical. Published DFT calculations have predicted a long list of candidates for 2D magnets in transition metal chalcogenides, transition metal halides, and other chemical families. Some calculations predict spectacularly high critical temperatures, for example: strained $NiCl_3$ ($T_c$~400 K) [130], $MnCl_3$ ($T_c$ = 680 K) [131], $Cr_3C_2$ ($T_c$ = 886 K) [132], $Mn_2NF_2$ ($T_c$ = 1877 K) [133], $Mn_2CF_2$ ($T_c$=520 K) [134], $Co_2S_2$ ($T_c$ =404 K) [35, 135]. Recently, magnetoelectric and multiferroic 2D magnets such as CrN have also been predicted [136]. For a thorough review of the theoretical predictions see Reference [35] and the references therein. Experimental validation of these predictions is a crucial missing step to test the accuracy of the latter methodology. As the standard form of local DFT does not include thermal fluctuations and dipolar interactions (shape anisotropy), such calculations must usually be combined with other techniques to predict magnetic stability. Currently there is no standard

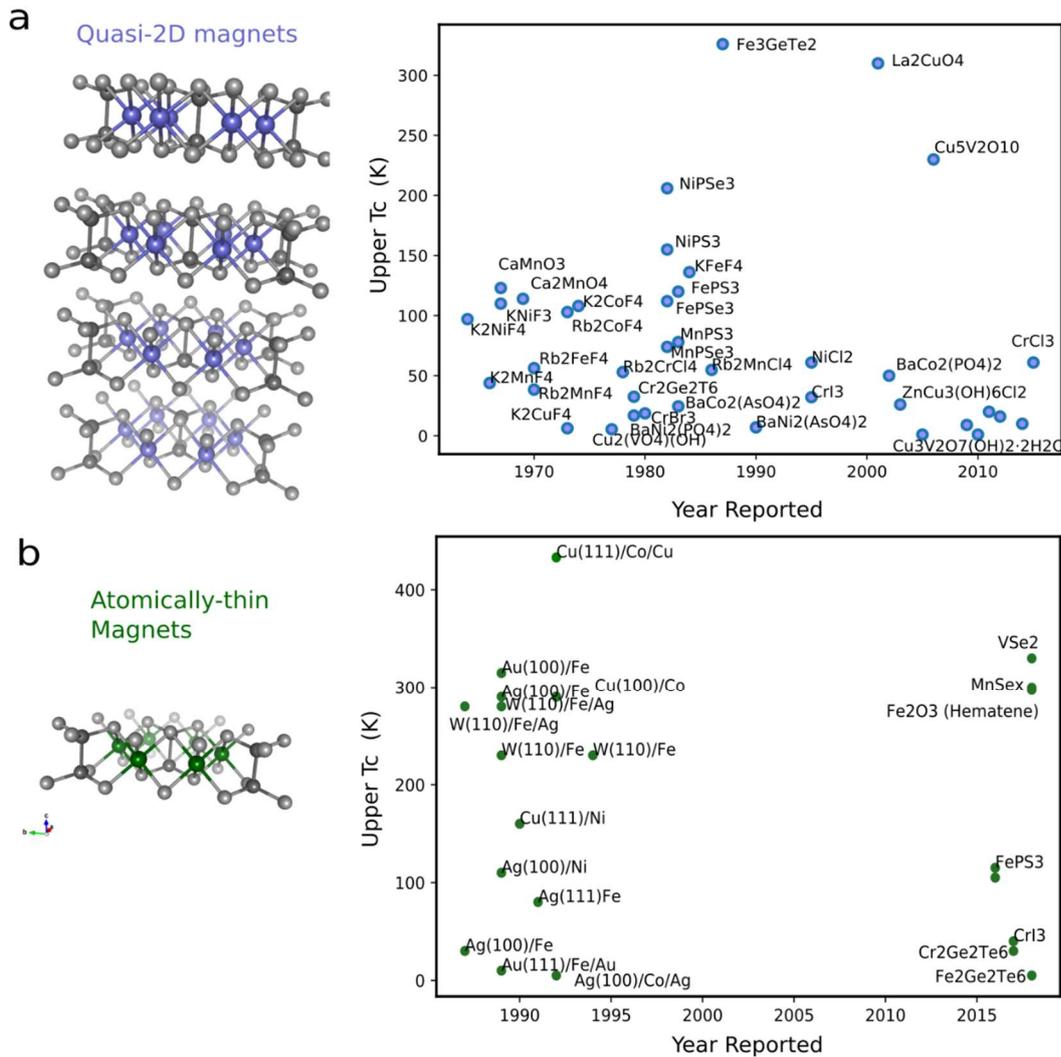

**Figure 7** a) Experimentally-known quasi-2D layered bulk crystals and their critical temperatures. a) Currently experimentally known two dimensional magnets and their critical temperatures. In cases where exact values are unknown, the upper limit is given based on past experimentsb theoretical approach to predict 2D magnets. Improved theoretical functionals and standardizing the prediction methods is key to ensure viable predictions of 2D magnets. In particular, the energy scale of magnetocrystalline anisotropy, which ultimately determines whether the Mermin-Wagner theorem applies is often on the scale of μeV which is challenging to predict accurately with *ab initio* calculations. Thus, rigorously predicting the thermodynamic stability of both the crystallographic and magnetic sub-systems in 2D magnets remains a lofty goal which will require further theoretical developments in the field of data-driven material design.

## 5. Conclusions and prospects

Although this review article covers the main three strands of 2D magnetism, it is not an exhaustive list of past work. 2D magnetism has a long and rich history intertwined with the development of modern physics and materials engineering, and it is simply impossible to capture all of the existing 2D magnets. For example, the idea of 2D magnetism in diluted semiconductors such as GaAs, GaN, ZnO, AlN, SiC, MoS2 and oxides has been explored [137-138] [139]. Recently, the exfoliation of a non-vdW material ($Fe_2O_3$) was reported, along with room temperature order [140]. In certain cases, interfaces of between two disparate materials can also become magnetic, which may be considered as another class of 2D magnet [85] [86]. It is therefore anticipated that many new low dimensional-magnets outside of the vdW family will be unveiled in the coming years.

Nevertheless, the immediate goal of achieving functional devices using 2D magnetic monolayers based on vdW compounds is now tantalizingly close. The key remaining challenge is to enhance the magnetic phase transition temperatures to above room temperature. While most recent work has focused only on the importance of the spin anisotropy for achieving this, historically two other factors have been identified as a way to overcoming the Mermin-Wagner theorem: long-range spin-spin interactions and shape anisotropy. The relative efficiency of these two schemes still needs to be assessed in real materials. An equally important challenge is to identify materials that are air-stable under ambient conditions, and sufficiently robust mechanically to allow for prolonged operation. While there have been several exhaustive theoretical explorations to identify 2D magnets with higher transition temperatures, the understanding of the mechanical and chemical stability is still in its infancy.

The current research climate has placed an intense focus on ferromagnets, and on 2D sheets that can be exfoliated from their 3D parent compounds. Within the new field of vdW magnets, there is currently a very strong emphasis on ferromagnetic $D=2$, $n=1,2,3$ systems, however, as a rule of thumb, the most exotic magnetic states occur at the extremes, for example, high $D$, low n or vice versa. These are mostly characterized as $D=2$, $n=1$ systems. While history has shown that most functional magnetic materials are ferromagnets, there are some precedents that show that complex order-parameters (eg. skyrmions, antiferromagnets) and short-range ordered magnets also have the potential to fill many important technological applications. Certain properties such as exchange-bias, multiferroic behavior, magnetoelectricity and current-induced domain wall motion depend on these more complex magnetic phases. Although single phase ferroelectrics and ferromagnetic 2D materials are now well-established, it is of great interest to test predictions of magnetoelectric and multiferroic 2D materials which combine these functionalities via the interplay of order parameters to potentially enable further functionality [136, 141]. It will be therefore valuable to extend the repertoire of experimental systems beyond the small number of current examples. A particularly interesting and unexplored fundamental case is $D=1$, $n=3$ of which there are potential examples in the vdW family. One implication is the possibility of forming spin ladders and spin chains (D=1) by using carefully selected materials and then performing exfoliation. In practice, achieving a perfect spin chain in a real crystal is extremely challenging because interplanar and interladder coupling is always present. More generally, the vdW magnets now offer a platform to explore low-dimensional magnetism (down to 1D and perhaps lower) to achieve exotic states of matter. By exfoliating an intrinsic vdW $D=1$ magnet, (eg. $AgCrP_2S_6$), it will be possible to design improved spin ladder systems by further suppressing interactions in one direction normal to the cleavage plane. It is expected that vdW atomically-thin magnets can reach $D<=2$ states where many of the interactions in the quasi-low dimensional crystal are strongly suppressed in the exfoliated layer, thus yielding magnets that exhibit more pure forms of quantum behavior.

Although recent work has focused on the spintronic applications of vdW materials, it is possible to envision that they could also operate as superb sensors in chemistry and biology. Ultra-thin films have been used in this capacity for some time, and the high-surface area of the vdW magnets is potentially advantageous. On-chip applications combining the magnetic and photonic functionality of transparent materials is an attractive option. Wafer-scale growth is an important task for ultimately enabling large scale circuitry, and novel synthesis methods are needed for this purposes. Another exciting prospective application is the potential for applications that move current whilst generating little or no heat, for example via the Quantum

Anomalous Hall effect. Such materials are expected to be dissipationless and lead to a pathway towards energy-efficient electronics. It has been proposed that this will require a 2D magnet which is also a topological insulator or Chern insulator. Currently, realizations in doped $Sb_2Te_3$ have Curie temperatures that are much too low, because the magnetism originated from dilute-magnetic semiconducting behavior[35]. It would be of great value to identify an intrinsic Chern insulator with a high Curie temperature in the vdW family. Many of the 2D magnets also Dirac spin-gapless semiconductors[142] or half metals with high spin polarization at the Fermi level[130], which holds great potential for realizing new forms of Anomalous Hall and Spin Hall effects[143]. Large high quality growth of low-dimensional magnets would then enable integration with large-scale spintronics circuitry. In the longer term, quantum low-dimensional magnets interfaces with superconductors, and the realization of 1D magnetic systems may also play a role in quantum computation and communication and as a playground for exotic quasiparticles. The next decade is rich with opportunity to study and functionalize low dimensional vdW magnets.

**Acknowledgements**
This research was partially supported by the Australian Research Council Centre of Excellence in Future Low-Energy Electronics Technologies (project number CE170100039) and funded by the Australian Government.

Received: ((will be filled in by the editorial staff))
Revised: ((will be filled in by the editorial staff))
Published online: ((will be filled in by the editorial staff))

**References**

**Table of contents entry:**

This review discusses two-dimensional magnets with applications for low-energy spintronics with an emphasis on experimental phenomena. Historical examples are discussed beginning in the 1960's. Recent progress using cleavable van der Waals materials in the period 2016-2018 is highlighted. The theoretical mechanisms for improved magnetic performance are introduced and discussed against the background of experimental evidence.

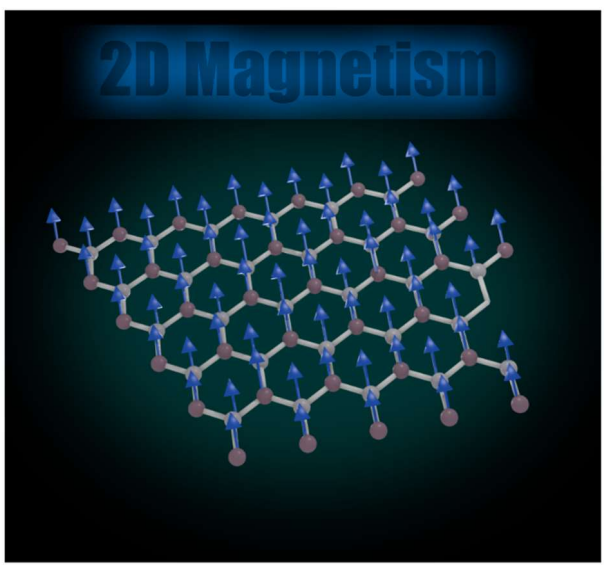

**Keyword**  magnetism

D. L. Cortie*, G. L. Causer, K. C. Rule, H. Fritzsche, W. Kreuzpaintner, F. Klose*

**Title** Two dimensional magnets: Forgotten history and recent progress towards spintronic applications

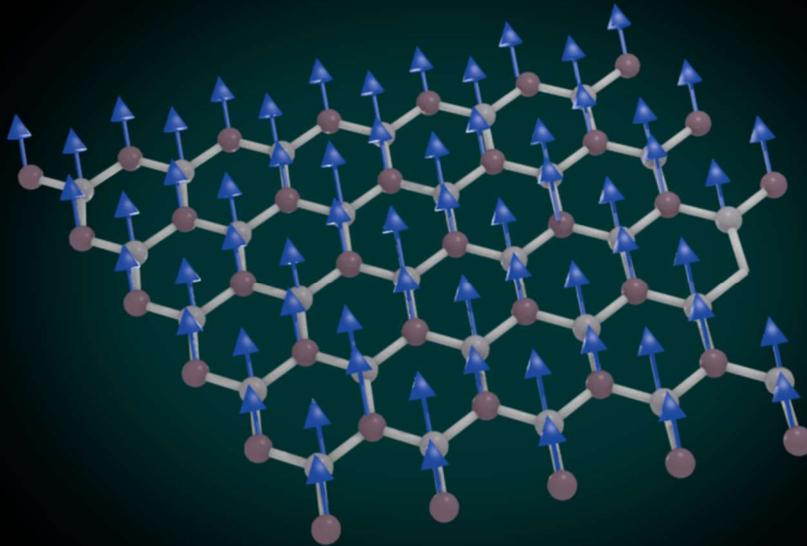

[1]     L. Onsager, *Phys. Rev.* **1944**, 65, 117.